\documentclass{rmf-d}
\usepackage{nopageno,rmfbib,multicol,times,epsf,amsmath,amssymb,cite}
\usepackage[latin1]{inputenc}
\usepackage[]{caption2}
\usepackage{graphics}

\usepackage{graphicx}
\usepackage[hidelinks]{hyperref}
\hypersetup{colorlinks, citecolor=black, linkcolor=black ,urlcolor={blue}}
\usepackage{wrapfig, blindtext}
\usepackage{comment}
\usepackage{tabularx}
\usepackage{float}

\usepackage{dblfnote}

\newcommand{\N}{\mathbb{N}}
\newcommand{\R}{\mathbb{R}}
\newcommand{\C}{{\kern+.25em\sf{C}\kern-.45em\sf{{\small{I}}} \kern+.45em\kern-.25em}}
\newcommand{\aso}{\hat{=}}
\newcommand{\be}{\begin{equation}}
\newcommand{\ee}{\end{equation}}
\newcommand{\bea}{\begin{eqnarray}}
\newcommand{\eea}{\end{eqnarray}}
\newcommand{\nn}{\nonumber}
\newcommand{\vep}{\varepsilon}
\def\rmfcornisa{Education of Physics} 

\def\rmfcintilla{{\it Rev.\ Mex.\ Fis.\/} } 
\clearpage \rmfcaptionstyle 
\begin{document}
\title{Ramanujan summation and the Casimir effect 
\vspace{-6pt}}
\author{Wolfgang Bietenholz}
\address{Instituto de Ciencias Nucleares, Universidad Nacional Aut\'{o}noma
de M\'{e}xico \\ A.P.\ 70-543, C.P.\ 04510 Ciudad de M\'{e}xico, M\'{e}xico}
\maketitle
\begin{abstract}
  \vspace{1em}
Srinivasa Ramanujan was a great self-taught Indian mathematician,
who died a century ago, at the age of only 32, one year after returning
from England. Among his numerous achievements is the
assignment of sensible, finite values to divergent series, which correspond
to Riemann's $\zeta$-function with negative integer arguments.
He hardly left any explanation about it, but following the few hints that
he gave, we construct a direct justification for the best known example,
based on analytic continuation.
As a physical application of Ramanujan summation we discuss the Casimir
effect, where this way of removing a divergent term corresponds to the
renormalization of the vacuum energy density, in particular of the photon
field. This leads to the prediction of the Casimir force between
conducting plates, which has now been accurately confirmed by experiments.
Finally we review the discussion about the meaning and interpretation of
the Casimir effect. This takes us to the mystery surrounding the magnitude
of Dark Energy.
  \vspace{1em}
\end{abstract}
\keys{Ramanujan summation; Casimir effect; renormalization; $\zeta$-function;
  Dark Energy
  \vspace{-4pt}}
\pacs{03.70.+k; 11.10.Gh \vspace{-4pt}}
\begin{multicols}{2}

We present some remarks about Ramanujan summation, the Riemann
$\zeta$-function and its application to the Casimir effect, following
the lines of our detailed review \cite{WB}.

\section{Ramanujan summation}

Srinivasa Ramanujan was a great Indian genius of mathematics,
an autodidact, who discovered numerous amazing formulae. He was born in
1887 in Erode and grew up in Kumbakonam, two towns in Eastern
India.\footnote{I thank T.R.\ Govindarajan for precise information;
  this was not accurate in Ref.\ \cite{WB}.}
\begin{figure}[H]
\centering
\includegraphics[scale=0.13]{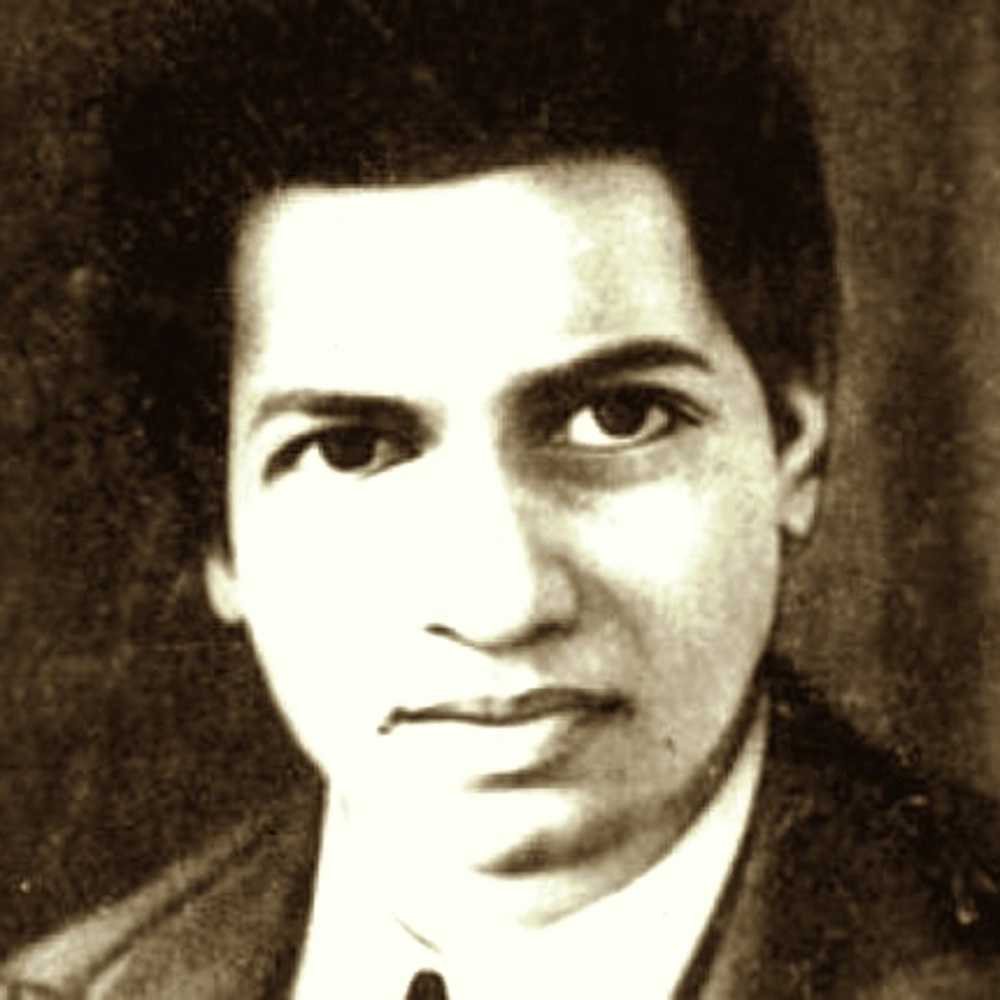}
\caption{Srinivasa Ramanujan (1887--1920)}
\label{Rama}
\vspace*{-3mm}
\end{figure}
After finishing High School he moved to Madras (today Chennai), where
he lived in extreme poverty. He was not admitted to university,
but he started to elaborate stunning mathematical formulae, 
often dealing with series.
Impressed by his achievements, the prominent British mathematician
Godfrey Hardy invited him to Cambridge UK from 1914--19. Hardy described
Ramanujan's discoveries as a ``process of mingled argument, and
intuition''.

An example is the following sequence of approximations to the number $\pi$,
\be
\frac{1}{\pi_{N}} \ = \ \frac{\sqrt{8}}{99^{2}} \ \sum_{n = 0}^{N}
\frac{(4n)!}{(4^{n} n!)^{4}} \ \frac{1103 + 26390 n}{99^{4n}} \ , 
\ee
with an extremely fast convergence: for $N= 0,1,2 \dots$
one obtains $| \pi - \pi_{N}| = {\cal O}(10^{-8(N+1)})$.

One of Ramanujan's famous achievements, which caused --- and still causes
--- confusion, was the assignment of finite values to divergent series.
Here are three examples, which he wrote down in a notebook \cite{Notebook},
and also in the first letter that he sent to Hardy when he still lived
in Madras,\footnote{We write $\aso$ for ``associated with'', whereas
  Ramanujan shocked people by writing a straight equal sign.}
\be  \label{Rsums}
\sum_{n = 1}^{\infty} 1 \ \aso \ -\frac{1}{2} \ , \ \
{\cal R} := \sum_{n=1}^{\infty} n \ \aso \ -\frac{1}{12} \ , \ \
\sum_{n = 1}^{\infty} n^{3} \ \aso \ \frac{1}{120} \ . \ \
\ee
This may look strange, but Hardy recognized the values of
Riemann's $\zeta$-function. For ${\rm Re} \, z > 1$ it is given by
\be
\zeta (z) = \sum_{n = 1}^{\infty} \frac{1}{n^{z}} \ .
\ee
Back in the 19th century, Bernhard Riemann had performed the analytic
continuation to $z \in \C - \{1\}$ \cite{Riemannzeta}. It can be
summarized with the identity
\be
\zeta (z) = \frac{(2\pi)^{z}}{\pi} \sin \left( \frac{\pi z}{2} \right)
\Gamma (1-z) \zeta (1-z) \ ,
\ee
which implies in particular
\be  \label{zetamk}
\zeta (-k) = - \frac{B_{k+1}}{k+1} \ , \quad k \in \N \ ,
\ee
which coincides with eqs.\ (\ref{Rsums}) for $k=0,1,2$.
$B_{k+1}$ are the Bernoulli numbers
(with the convention $B_{1} = 1/2)$.

Best known is the case of
${\cal R} \ \aso \ \zeta(-1)$, where Ramanujan only documented two
intermediate steps \cite{Notebook},
\bea
{\cal E} \!\!\! &:=&  \!\!\! 1-2+3-4+5 \dots \ \aso \ \frac{1}{4} \nn \\
{\cal R} - {\cal E}  \!\!\! & \aso &  \!\!\! 4 + 8 + 12 + \dots \ \aso \
4 {\cal R} \ \ \Rightarrow \ \ {\cal R} \ \aso \ -\frac{1}{12} \ . \quad
\label{ERseries}
\eea
This looks like uncontrolled operations on divergent series,
but the sensible results --- in three cases, cf.\ eqs.\ (\ref{Rsums})
--- cannot be by accident.
Ramanujan mostly worked on a slate, he wrote down only little on paper,
which was expensive. We can only speculate about his undocumented
intermediate steps. Here we present a reasoning, which follows the lines
of eqs.\ (\ref{ERseries}) and clarifies the meaning of ${\cal R}$.
It invokes analytic continuation, which is the only valid justification.

Let us consider $|z|<1$, and the limit $z \to -1$, for
\begin{eqnarray}
G(z) \!\!\!&:=&\!\!\! \sum_{n = 0}^{\infty} z^{n} = \frac{1}{1-z}
\overbrace{\longrightarrow}^{z \to -1}
\overbrace{1-1+1-1+ \dots}^{\mbox{\small{Grandi's series}}} \ \aso \ \frac{1}{2}
\nn \\
G'(z) \!\!\!&=&\!\!\! \sum_{n=0}^{\infty} nz^{n-1} = \frac{1}{(1-z)^{2}}
\ \overbrace{\longrightarrow}^{z \to -1} \ {\cal E} \ \aso \ \frac{1}{4} \ .
\end{eqnarray}
The original series (on the left) only converge for $|z|<1$, but we
arrive at holomorphic (complex analytic) functions, with unique analytic
continuations to $\C - \{ 1 \}$.
$G(z)$ is the familiar geometrical series, and in its derivative $G'(z)$
the series ${\cal R}$ requires $z\to 1$, which is singular.
We regularize ${\cal R}$ in two more ways, again at $|z|<1$,
\begin{eqnarray*}
  R_{1}(z) \!\!\!&=&\!\!\! 1 -2z + 3z^{2} - 4 z^{3} \dots
  = \frac{1}{(1+z)^{2}} \ , \\
  R_{2}(z) \!\!\!&=&\!\!\! 1 +2z^{2} + 3z^{4} + 4 z^{6} \dots
  = \frac{1}{(1-z^{2})^{2}} \ ,
\end{eqnarray*}
with divergent limits $R_{1,2}(z \to -1)$. However, in the linear combination
$$
R_{1}(z) + 4z R_{2}(z) = G'(z)
$$
the poles in the Laurent series of $z = -1 + \vep$ cancel. We obtain
$\tfrac{1}{4} + {\cal O}(\vep )$, so now the analytic continuation
to $z=-1$ works,
\be
{\cal R} - 4 {\cal R}  \ \aso \ \frac{1}{4} \quad \Rightarrow \quad
{\cal R} \ \aso \ -\frac{1}{12} \ . 
\ee

In Ref.\ \cite{WB} we also derived $\zeta (0) = -1/2$ in a similar way.
For a general and systematic discussion of Ramanujan summation we refer
to Ref.\ \cite{Candelpergher}.

Is this just mathematical entertainment? No, it applies to quantum field
theory: for a suitable system, Ramanujan summation removes a counter-term
in a physically sensible manner, and provides renormalized results.
In particular, it predicts a force, which has now been experimentally
measured, but let us begin with a toy model.

\section{The 1d Casimir effect}

The Casimir effect was first predicted in Ref.\ \cite{CasimirPolder},
but here we follow the point of view which Hendrik Casimir (1909--2000)
expressed a little later \cite{Casimir48}, after a discussion
with Niels Bohr. For comprehensive overviews, we refer to
Refs.\ \cite{Milton,Oxbook}.

\begin{figure}[H]
\vspace{-5.5cm}
\centering
\includegraphics[scale=0.5]{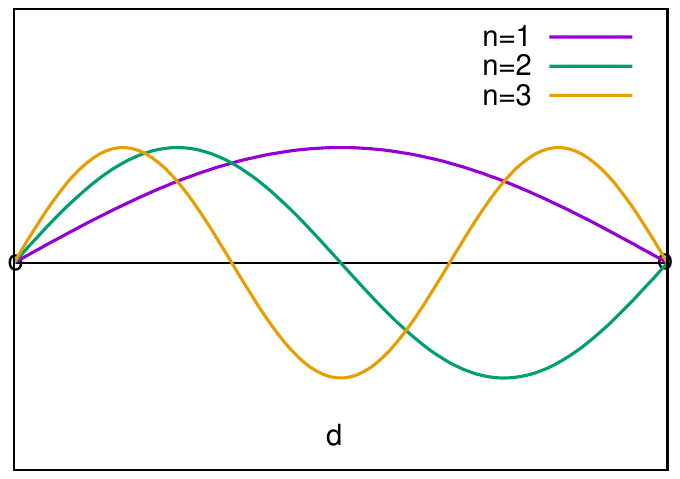}
\vspace{-1.5cm}
\caption{A symbolic illustration of the
  leading standing waves between two Dirichlet boundaries in one
  spatial dimension, separated by some distance $d$.}
\label{1dCasimir}
\end{figure}

As a toy model, we first consider a free, massless, neutral scalar field
in one spatial dimension, $\phi (t,x) \in \R$.
We impose Dirichlet boundaries at $x=0$ and $d$,
which enforce $\phi(t,0) = \phi(t,d) =0$. Inside this interval, the
configurations can be expanded in terms of standing waves, as
symbolically illustrated
in Figure \ref{1dCasimir}, with wave numbers $k_{n} = \frac{n\pi}{d}$,
$n=1,2,3 \dots$, and ground state energies $E_{n} = \tfrac{1}{2} k_{n}$ \
(in natural units, $\hbar = c = 1$).

\noindent
In the interval $[\, 0,\, d \, ]$ we obtain the bare ($\rho$) and the
renormalized ($\rho_{\rm r}$) vacuum energy density,
\be  \label{rho1d}
\rho (d) = \frac{1}{2d} \sum_{n\geq 1} k_{n}
= \frac{\pi}{2d^{2}} \sum_{n\geq 1} n \ \aso \ - \frac{\pi}{24 d^{2}}
= \rho_{\rm r}(d) \ .
\ee
Hence the renormalized energy between the boundaries amounts
to $E_{\rm r} (d) = d \, \rho_{\rm r}(d)$.
If we consider the boundary at $d$ as flexible,
we obtain the force $F_{\rm r}(d)$ acting on it,
\be  \label{1dforce}
F_{\rm r}(d) = -E_{\rm r}'(d) = - \frac{\pi}{24d^{2}} \ .
\ee
We see that the boundaries are {\em attracted} to each other.

In the last step of eq.\ (\ref{rho1d}) we have applied Ramanujan
summation: it amounts to subtracting the {\em counter-term} $\rho (\infty )$,
which provides the {\em renormalized} energy density and force.

To confirm this property, notice that for $\rho (\infty )$ the sum over
the modes $k_{n}$ turns into an integral. Hence the renormalized
energy density $\rho_{\rm r}(d)$ emerges as the difference between a sum
and an integral, which can be expanded with the Euler-Maclaurin formula,
see {\it e.g.}\ Ref.\ \cite{AS}. Ref.\ \cite{WB} discusses
in detail how this leads to the result in eq.\ (\ref{rho1d}) ---
in agreement with Ramanujan summation, in particular with
${\cal R} \ \aso \ -1/12$, the famous result that we re-derived in Section 1.

The application of the Euler-Maclaurin formula requires some regularization
(with several ingredients), but the final result confirms eq.\
(\ref{zetamk}), and therefore also (\ref{Rsums})
and (\protect\ref{rho1d}), for any choice of the regularization
(if certain conditions are fulfilled);
this is a generic feature of renormalization.\footnote{Of course,
  in the case of interacting fields, renormalization takes more
  than subtracting a divergent term, see {\it e.g.}\ Ref.\ \cite{Collins}.
  In general one assigns renormalized, energy-dependent values to the
  fields and their couplings.}

If we want to safely exclude effects from the vacuum energy {\em outside}
the given interval, we can assume three Dirichlet boundaries, where only
the central one (the ``piston'') is flexible. This is also
discussed in Ref.\ \cite{WB}, and it justifies the form of the
renormalized force in eq.\ (\ref{1dforce}).

\section{3d Casimir effect of a photon field}

\begin{figure}[H]
\centering
\includegraphics[scale=0.45]{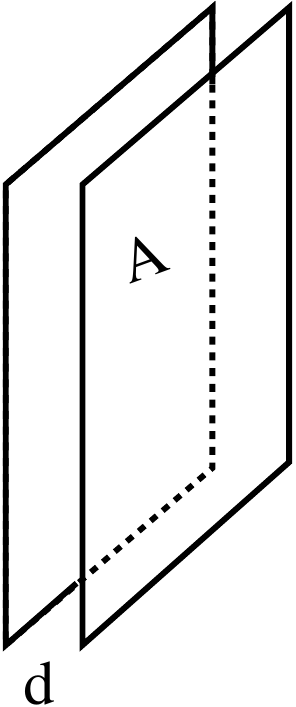}
\caption{Set-up for the phenomenological
  Casimir effect between two conducting plates.
  For this setting, the Casimir force was first successfully
  measured in an experiment in Padua \cite{Padua}.}
\label{3dCasimir}
\end{figure}

Let us proceed to a realistic setting in 3 spatial dimensions: it consists
of two parallel plates, which are (perfectly or at least well) conducting,
both with area $A$, separated by a short distance $d \ll \sqrt{A}$, as
sketched in Figure \ref{3dCasimir}.

It is a good approximation to consider only the vacuum energy {\em between}
the plates, $E(d) = A d \rho(d) $, obtained from the photon ground
state energy density $\rho (d)$. In the corresponding formula, we treat
the momentum components parallel to the plates, $k_{1},\, k_{2}$,
as continuous, while the vertical component is discrete,
as in the 1d case of Section 2,
\bea
E(d) &=& \frac{A}{(2 \pi)^{2}} \int dk_{1} dk_{2}
\sum_{n = 0}^{\infty} \sqrt{k_{1}^{2} + k_{2}^{2} + (\tfrac{\pi n}{d})^{2} } \nn \\
&=& \frac{A}{6\pi} \sum_{n = 0}^{\infty} \left.
\Big[ K^{2} + (\tfrac{\pi n}{d})^{2} \Big]
^{3/2} \right|^{\infty}_{K=0}
\eea
(a factor of 2 accounts for the photon polarizations).
Infinitesimally short waves are not sensitive to the presence
of the plates (at finite separation $d$). Therefore only the
lower bound, $K=0$, contributes to the difference
$\rho_{\rm r}(d) = \rho(d) - \rho(\infty )$, which leads to
\bea
&& \hspace*{-8mm}
\frac{E_{\rm r}(d)}{A} \aso - \frac{1}{6\pi} \frac{\pi^{3}}{d^{3}}
\zeta (-3) = - \frac{\pi^{2}}{720 d^{3}} \ ,
\quad F_{\rm r}(d) = -E_{\rm r}'(d)
\nn \\ && \hspace*{-8mm}
\hspace*{10mm} = \, -\frac{\pi^{2}A}{240 d^{4}}
\simeq - 1.3  \cdot 10^{-7} \ {\rm N} \,
\Big( \frac{\mu {\rm m}}{d} \Big)^{4}
\frac{A}{{\rm cm}^{2}} \ ,
\label{3dforce}
\eea
where we applied the last Ramanujan sum of eq.\ (\ref{Rsums}).
To arrive at a force in terms of Newton (N), we had to insert a
factor $\hbar c$, which indicates that we are
dealing with a relativistic quantum effect.

Again, Ramanujan summation provides a renormalized result,
and again the force is {\rm attractive,} although the signs of
$\zeta (-1)$ and $\zeta (-3)$ are opposite (the lower integral bound
at $K=0$ flips the sign once more).

This force was conclusively measured, first in 1997/8, with
about 5\% accuracy \cite{Lamoreaux,MohideenRoy}. In these experiments,
the geometry was a plate and a sphere, because it is very difficult to
keep two plates exactly parallel. The first successful experiment
with parallel plates, as in Figure 3, was performed by a collaboration
in Padua, Italy \cite{Padua}. They used parallel silicon stripes
with area $A=1.9 \, {\rm cm} \times 1.2 \, {\rm mm}$, and their
separation varied from $d=0.5 \, \mu {\rm m}$ to $ 3 \,
\mu {\rm m}$. According to eq.\ (\ref{3dforce}) this implies an
attractive Casimir force of strength
$F_{\rm r} = -4.7 \cdot 10^{-7}\ {\rm N}$ to $-3.7 \cdot 10^{-10}\ {\rm N}$.
This force could be measured by using a fiber-optic interferometer,
and monitoring the shift in the resonator frequency.

\vspace*{-2mm}
\section{Interpretation of the Casimir Effect}

What kind of force is this?
At first sight, it does not seem to be part of the interactions
described by the Standard Model. However, since it refers to the
photon field, this effect must be electromagnetic, {\it i.e.}\
a facet of Quantum Electrodynamics (QED).\footnote{In
  principle it should exist for other fields as well, but if they
  are massive or self-interacting, the corresponding Casimir effect
  in realistic setting tends to be strongly suppressed or over-shadowed.}
What seems strange, however, is that the coupling $\alpha$ did
not appear in the considerations of Section 3.
We will come back to this puzzle in Subsection 4.1.

The Casimir force was derived from the QED vacuum energy density, a quantity,
which is not directly observable. So does its experimental confirmation
imply indirect evidence for the physical existence of the QED vacuum energy?
This is a wide-spread point of view, which is expressed {\it e.g.}\ in
the books \cite{Milton,Oxbook}, and the review \cite{SaSta}
even writes that the existence of the vacuum energy density of the
photon field ``has been spectacularly demonstrated by the Casimir effect''.

Actually a direct manifestation of the vacuum energy density
in the Universe does exist. Here we refer to Dark Energy, which
(essentially) corresponds to the Cosmological Constant. It provides the
most natural explanation for the accelerated expansion of the Universe,
which was observed at the end of the 20th century, and which corresponds
to $\rho_{\rm DE} \approx 2 \cdot 10^{-3}\ {\rm eV}^{4}$.

One is tempted to relate it to $\rho (\infty )$, with a cutoff,
most naturally at the Planck scale,
$E_{\rm Planck} \simeq 1.2 \cdot 10^{28}~{\rm eV}$,
but the density obtained is this manner is {\em much} larger
than the Dark Energy density.
If we truncate the momentum integral at the Planck scale,
\be
\rho = \frac{1}{(2\pi )^{3}} \int_{|k| \leq E_{\rm Planck}} d^{3}k \ k
= \frac{1}{8\pi^{2}} E_{\rm Planck}^{4} \ ,
\ee
we obtain a prediction for the vacuum energy, which is about
121 orders of magnitude too large
--- perhaps the worst prediction in the history of science ---
and the reason for this fiasco is still not well understood
(a pedagogical discussion is given in the appendix of Ref.\ \cite{WBp}).

In a world with unbroken supersymmetry the (positive) bosonic and
(negative) fermionic vacuum energy cancel.
However, even if one still assumes supersymmetry to exist, it has to be
broken in our low-energy world, such that this cancellation
takes place only in part. One still obtains a vacuum energy, which is
at least 60 orders of magnitude too large \cite{Carroll}, so supersymmetry
does not solve the problem.

Just for fun, let us consider the Dark Energy density,
which corresponds to the observation, $\rho_{\rm DE}$,
and ask what cavity between two conducting plates it would take
to obtain the same renormalized vacuum energy density, according
to eq.\ (\ref{3dforce}). Actually the sign is different, so let
us consider the absolute value.
The condition $\rho_{\rm DE} = |\rho_{\rm r}(d)| = \pi^{2}/ (720 d^{4})$
requires the distance $d \approx 0.3\, \mu{\rm m}$, which happens
to be close to the minimal separation in the Padua experiment.

\subsection{Where does the electromagnetic coupling $\alpha$ appear?}

Now let us address the r\^{o}le of the electromagnetic coupling
$\alpha$, as we promised. It was analyzed in particular by
Jaffe {\it et al.}; their point of view is summarized in Ref.\ \cite{RJaffe}.
They insist that the original picture by Casimir and Polder
\cite{CasimirPolder} was the correct one, {\it i.e.}\ they consider
the Casimir force as a pure {\em van der Waals force.}\footnote{We mean
  the van der Waals --- or London--van der Waals --- force in the
  narrow sense: a collective, induced, (usually) attractive multi-pole
  interaction between (electrically neutral) molecules or atoms.}
Jaffe concludes
that ``Casimir forces can be computed without reference to zero-point
energy'', which would mean that their observation does {\em not}
imply the physical existence of the QED vacuum energy.

Jaffe {\it et al.}\ do obtain an $\alpha$-dependent Casimir force,
$F_{\rm r}(\alpha )$, with $F_{\rm r}(\alpha =0)=0$. This seems trivial,
but their result at $\alpha \to \infty$ is amazing: they derive a
finite Casimir force in this limit, which coincides with $F_{\rm r}$
in eq.\ (\ref{3dforce}), as obtained from the
vacuum energy consideration. This explains why $\alpha$ did not
appear there, but then we are left with the question why this
result matches the observations so well. It turns out that in an
experimentally realistic setting, the exact force is close to
this approximation \cite{RJaffe},
\be
F_{\rm r}(\alpha \gg 10^{-5}) \simeq F_{\rm r}(\alpha = \infty)
= - \frac{\pi^{2}}{240d^{4}} \ ,
\ee
which easily holds for $\alpha \simeq 1/137$.
This picture seems consistent, so is the effect de-mystified?

\subsection{Are there repulsive Casimir forces?}

There are objections against the latter point of view, which are expressed
{\it e.g.}\ by Lamoreaux \cite{Lamoreaux}. He insists that Casimir force
and van der Waals force are conceptually different, because
``the van der Waals force is always attractive, whereas the sign
of the Casimir force is geometry dependent.''

Indeed, numerous theoretical studies predict a repulsive Casimir
force for certain geometric settings \cite{Boyer}, such as specific
parallelepipeds \cite{Jan}. Jaffe {\it et al.}\
reject the repulsive scenario, and here the discussion enters
subtle details \cite{HJKS}.
Ref.\ \cite{Milton} approves the equivalence of van der Waals and
Casimir forces, but insists that they can still be repulsive.
  
A repulsive Casimir force was actually measured in 2009 \cite{Munday},
in agreement with a historic prediction in Ref.\ \cite{DLP},
but for materials immersed in a fluid. Hence the physical existence of
$\rho_{\rm QED}$ --- in vacuum --- remains an exciting, open question.

Since particle physics is very well described by Quantum Field
Theory, one might argue that quantum fields --- and therefore
also the QED vacuum --- are practically inevitable. However, the
Casimir force was also derived with a Green's function technique,
in the framework of QED, {\em without} referring to the QED vacuum
energy density \cite{Schwinger}.

Of course, the question whether or not $\rho_{\rm QED}$ exists
is usually irrelevant --- in general we only care about energy
{\em differences}, so an additive constant does not matter.
An exception is the expansion of the Universe, and {\em perhaps}
--- depending on its interpretation --- the Casimir effect.

\appendix

\section{A photon in a Casimir cavity}

An interesting prediction is the {\em Scharnhorst effect:} it states that
in a Casimir cavity between parallel plates, the speed of light in
vacuum --- for photons travelling perpendicular to the plates ---
should increase \cite{Scharnhorst}. The predicted effect
is so tiny that it cannot be experimentally tested --- for instance
for a plate separation of 1~$\mu$m the relative increase would be
of ${\cal O}(10^{-32})$. Still it led to a discussion if this could
--- in principle --- lead to a causality paradox. This issue is
reviewed and extensively discussed in Ref.\ \cite{Clark}.

We end with a simpler question: does a photon, which passes through a Casimir
cavity (for instance parallel to the plates) change its energy, in the
spirit of Bernoulli's Principle in fluid dynamics? \\

\noindent
{\bf Acknowledgments:} I would like to thank the organizers of the
{\it XXXV Reuni\'{o}n Anual de la Divisi\'{o}n de Part\'{\i}culas y Campos}
of the {\it Socieded Mexicana de F\'{\i}sica}
where this work was presented. I further thank Barnab\'{a}s Deme and
Kimball Milton for helpful communication. This work was supported by
UNAM-DGAPA-PAPIIT, grant number IG100219.
    
\end{multicols}
\medline
\begin{multicols}{2}

\end{multicols}

\begin{thebibliography}{99}

  \bibitem{WB} W.\ Bietenholz,
From Ramanujan to renormalization: the art of doing
away with divergences and arriving at physical results,
{\it Rev.\ Mex.\ F\'{\i}s.\ E} {\bf 18} (2021) 020203
[arXiv:2102.09371 [physics.hist-ph]],
\href{https://doi.org/10.31349/RevMexFisE.18.020203}{10.31349/RevMexFisE.18.020203}
 
\bibitem{Notebook} S.\ Ramanujan, Second Notebook (unpublished),
  Chapter VI. \\
  \{reproduced in Chapter 8 of Ref.\ \cite{Candelpergher}\}

\bibitem{Riemannzeta} B.\ Riemann, \"{U}ber die Anzahl der Primzahlen
  unter einer gegebenen Gr\"{o}\ss e, in {\it Monatsberichte der
  Berliner Akademie,} November 1859.
  
\bibitem{Candelpergher} B.\ Candelpergher,
  Ramanujan summation of divergent series,
  Lectures Notes in Mathematics, 2185 [hal-01150208v2], 2017,
\href{https://hal.univ-cotedazur.fr/hal-01150208v2/document}{hal.univ-cotedazur.fr/hal-01150208v2/document}

\bibitem{CasimirPolder} H.\ B.\ G.\ Casimir and D.\ Polder,
  The Influence of Retardation on the London-van der Waals Forces,
  {\it Phys.\ Rev.}\ {\bf 73} (1948) 360--372, 
\href{https://doi.org/10.1103/PhysRev.73.360}{10.1103/PhysRev.73.360}
 
\bibitem{Casimir48} H.\ B.\ G.\ Casimir,
On the attraction between two perfectly conducting plates,
{\it Proc.\ Kon.\ Ned.\ Akad.\ Wetensch.\ B} {\bf 51} (1948) 793--795.

\bibitem{Milton} K.\ A.\ Milton, The Casimir effect: Physical
  manifestations of zero-point energy, World Scientific, 
  2001.

\bibitem{Oxbook} M.\ Bordag, G.\ L.\ Klimchitskaya, U.\ Mohideen and
V.\ M.\ Mostepanenko, Advances in the Casimir Effect,
Oxford Scholarship Online, 2009, \\
\href{https://DOI:10.1093/acprof:oso/9780199238743.001.0001}{10.1093/acprof:oso/9780199238743.001.0001}

\bibitem{Collins} J.\ C.\ Collins, Renormalization,
Cambridge University Press, 1984.

\bibitem{AS} M.\ Abramowitz and I.\ Stegun,
  Handbook of Mathematical Functions with Formulas, Graphs, and
  Mathematical Tables, Dover Publications, 1972 (10th edition).
\href{https://personal.math.ubc.ca/~cbm/aands/abramowitz\_and\_stegun.pdf}{personal.math.ubc.ca/$\sim$cbm/aands/abramowitz\_and\_stegun.pdf}

\bibitem{Padua} G.\ Bressi, G. \ Carugno, R.\ Onofrio and G.\ Ruoso,
Measurement of the Casimir Force between Parallel Metallic Surfaces,
{\it Phys.\ Rev.\ Lett.}\ {\bf 88} (2002) 041804
[quant-ph/0203002],
\href{https://doi.org/10.1103/PhysRevLett.88.041804}{10.1103/PhysRevLett.88.041804}

\bibitem{Lamoreaux} S.\ K.\ Lamoreaux,
Demonstration of the Casimir Force in the 0.6 to 6 $\mu$m Range,
{\it Phys.\ Rev.\ Lett.}\ {\bf 78} (1997) 5--8
\href{https://doi.org/10.1103/PhysRevLett.78.5}{10.1103/PhysRevLett.78.5}
[Erratum: {\it Phys.\ Rev.\ Lett.}\ {\bf 81} (1998) 5475--5476,
\href{https://doi.org/10.1103/PhysRevLett.81.5475}{10.1103/PhysRevLett.81.5475}]

\bibitem{MohideenRoy} U.\ Mohideen and A.\ Roy,
Precision Measurement of the Casimir Force from 0.1 to 0.9~$\mu$m,
{\it Phys.\ Rev.\ Lett.}\ {\bf 81} (1998) 4549
[physics/9805038],
\href{https://doi.org/10.1103/PhysRevLett.81.4549}{10.1103/PhysRevLett.81.4549}

\bibitem{SaSta} V.\ Sahni and A.\ Starobinsky,
The Case for a Positive Cosmological $\Lambda$-term,
{\it Int.\ J.\ Mod.\ Phys.\ D} {\bf 9} (2000) 373--443
[astro-ph/9904398],
\href{https://doi.org/10.1142/S0218271800000542}{10.1142/S0218271800000542}

 \bibitem{WBp} W.\ Bietenholz, What are Elementary Particles?
From Dark Energy to Quantum Field Excitations,
{\it Rev.\ Cub.\ F\'{\i}s.}\ {\bf 37} (2020) 146--151
[arXiv:2011.07719 [physics.pop-ph]],\\
\href{http://www.revistacubanadefisica.org/RCFextradata/OldFiles/2020/v37n2/RCF2020v37p146.pdf}{www.revistacubanadefisica.org/RCFextradata/OldFiles/2020/\\v37n2/RCF2020v37p146.pdf}

\bibitem{Carroll} S.\ M.\ Carroll, The Cosmological Constant,
{\it Living Rev.\ Rel.}\ {\bf 4:1} (2001) 1--56
[astro-ph/0004075],
\href{https://doi.org/10.12942/lrr-2001-1}{10.12942/lrr-2001-1}

\bibitem{RJaffe} R.\ L.\ Jaffe,
The Casimir Effect and the Quantum Vacuum,
{\it Phys.\ Rev.\ D} {\bf 72} (2005) 021301
[hep-th/0503158],
\href{https://doi.org/10.1103/PhysRevD.72.021301}{10.1103/PhysRevD.72.021301}

\bibitem{Boyer} T.\ H.\ Boyer,
  Quantum Electromagnetic Zero-Point Energy of a Conducting
  Spherical Shell and the Casimir Model for a Charged Particle,
  {\it Phys.\ Rev.}\ {\bf 174} (1968) 1764--1774,
\href{https://doi.org/10.1103/PhysRev.174.1764}{10.1103/PhysRev.174.1764}

\bibitem{Jan} S.\ G.\ Mamaev and N.\ N.\ Trunov,
Dependence of the vacuum expectation values of the energy-momentum
tensor on the geometry and topology of the manifold,
{\it Theor.\ Math.\ Phys.}\ {\bf 38} (1979) 228--234,
\href{https://doi.org/10.1007/BF01018540}{10.1007/BF01018540}.
J.\ Ambj{\o}rn and S.\ Wolfram,
Properties of the vacuum. I. Mechanical and thermodynamic,
{\it Annals Phys.}\ {\bf 147} (1983) 1--32,
\href{https://doi.org/10.1016/0003-4916(83)90065-9}{10.1016/0003-4916(83)90065-9}.
G.\ J.\ Maclay,
Analysis of zero-point electromagnetic energy and Casimir forces
in conducting rectangular cavities,
{\it Phys.\ Rev.\ A} {\bf 61} (2000) 052110,
\href{https://doi.org/10.1103/PhysRevA.61.052110}{10.1103/PhysRevA.61.052110}.
M.\ Bordag, U.\ Mohideen and V.\ M.\ Mostepanenko,
New Developments in the Casimir Effect,
{\it Phys.\ Rept.}\ {\bf 353} (2001) 1--205
[quant-ph/0106045],
\href{https://doi.org/10.1016/S0370-1573(01)00015-1}{10.1016/S0370-1573(01)00015-1}

\bibitem{HJKS} R.\ M.\ Cavalcanti,
Casimir force on a piston,
{\it Phys.\ Rev.\ D} {\bf 69} (2004) 065015
[quant-ph/0310184],
\href{https://doi.org/10.1103/PhysRevD.69.065015}{10.1103/PhysRevD.69.065015}.
M.\ P.\ Hertzberg, R.\ L.\ Jaffe, M.\ Kardar
and A.\ Scardicchio, Attractive Casimir Forces in a Closed Geometry,
{\it Phys.\ Rev.\ Lett.}\ {\bf 95} (2005) 250402
[quant-ph/0509071],
\href{https://doi.org/10.1103/PhysRevLett.95.250402}{10.1103/PhysRevLett.95.250402};
Casimir Forces in a Piston Geometry at Zero and Finite Temperatures,
{\it Phys.\ Rev.\ D} {\bf 76} (2007) 045016
[arXiv:0705.0139 [quant-ph]],
\href{https://doi.org/10.1103/PhysRevD.76.045016}{10.1103/PhysRevD.76.045016}

\bibitem{Munday} J.\ N.\ Munday, F.\ Capasso and V.\ A.\ Parsegian,
  Measured long-range repulsive Casimir-Lifshitz forces,
{\it Nature} {\bf 457} (2009) 170--173,
\href{https://doi.org/10.1038/nature07610}{10.1038/nature07610}

\bibitem{DLP} I.\ E.\ Dzyaloshinskii, E.\ M.\ Lifshitz and
  L.\ P.\ Pitaevskii, General theory of van der Waals' forces,
  {\it Sov.\ Phys.\ Uspekhi} {\bf 4} (1961) 153--176,
\href{https://doi.org/10.1070/PU1961v004n02ABEH003330}{10.1070/PU1961v004n02ABEH003330}

\bibitem{Schwinger} J.\ Schwinger, Casimir effect in source theory,
{\it Lett.\ Math.\ Phys.}\ {\bf 1} (1975) 43--47,
\href{https://doi.org/10.1007/BF00405585}{10.1007/BF00405585}.
J.\ Schwinger, J.\ De\-Raad and K.\ A.\ Milton,
Casimir Effect in Dielectrics,
{\it Ann.\ Phys.\ (N.Y.)} {\bf 115} (1978) 1--23,
\href{https://doi.org/10.1016/0003-4916(78)90172-0}{10.1016/0003-4916(78)90172-0}

\bibitem{Scharnhorst} K.\ Scharnhorst, On propagation of light in
  the vacuum between plates,
  {\it Phys.\ Lett.}\ {\bf B} 236 (1990) 354--359,
\href{https://doi.org/10.1016/0370-2693(90)90997-K}{10.1016/0370-2693(90)90997-K}
  [Erratum: Phys.\ Lett.\ B 787 (2018) 204,
\href{https://doi.org/10.1016/j.physletb.2018.12.002}{10.1016/j.physletb.2018.12.002}].
G.\ Barton, Faster-than-$c$ light between parallel mirrors.
The Scharnhorst effect rederived,
{\it Phys.\ Lett.\ B} {\bf 237} (1990) 559--562,
\href{https://doi.org/10.1016/0370-2693(90)91224-Y}{10.1016/0370-2693(90)91224-Y}

\bibitem{Clark} S.\ G.\ de Clark,
The Scharnhorst Effect: Superluminality and Causality
in Effective Field Theories,
Ph.D.\ thesis, University of Arizona, 2016,\\
\href{https://repository.arizona.edu/handle/10150/622964}{repository.arizona.edu/handle/10150/622964}


\end{thebibliography}
\end{document}